\documentclass[prl,showpacs,twocolumn]{revtex4}

\newcommand{\ket}[1]{|{#1}\rangle}

\newcommand{\be}{\begin{equation}}
\newcommand{\ee}{\end{equation}}

\usepackage{amsfonts}
\usepackage{amsmath}
\usepackage{graphicx}
\usepackage{amssymb}
\usepackage{amsmath}
\usepackage{amssymb}
\usepackage{graphicx}
\usepackage{lscape}
\usepackage{color}
\usepackage{epstopdf}

\begin{document}
\title{Flat band of topological states bound to a mobile impurity}

\author{M. Valiente}
\affiliation{Institute for Advanced Study, Tsinghua University, Beijing 100084, China}

\begin{abstract}
I consider a particle in the topologically non-trivial Su-Schrieffer-Heeger (SSH) model interacting strongly with a mobile impurity, whose quantum dynamics is described by a topologically trivial Hamiltonian. A particle in the SSH model admits a topological zero-energy edge mode when a hard boundary is placed at a given site of the chain, which may be modelled by a static impurity. By solving the two-body problem analytically I show that, when the impurity is mobile, the topological edge states of the Su-Schrieffer-Heeger model remain fully robust and a flat band of bound states at zero energy is formed as long as the continuum spectrum of the two-body problem remains gapped, without the need for any boundaries in the system. This is guaranteed for a sufficiently heavy impurity. As a consequence of the infinite degeneracy of the zero energy modes, it is possible to spatially localise the particle-impurity bound states, effectively making the impurity immobile. These effects can be readily observed using two-dimensional photonic lattices. 
\end{abstract}
\pacs{
}
\maketitle
With the theoretical proposal \cite{Bernevig,Kane} and subsequent experimental realisation \cite{Koenig} of quantum spin Hall insulators, which are protected by symmetry, as opposed to the usual integer \cite{Hall1,Hall2,Hall3} and fractional \cite{Fractional1,Fractional2} quantum Hall effects, the field of topologically non-trivial matter has experienced a major surge in research interest. Remarkably, it is not only in condensed matter systems that topological effects play a role, but also in ultracold atomic ensembles \cite{Cooper}, classical optical systems such as photonic lattices \cite{reviewphotonics}, and even mechanical metamaterials \cite{HuberSusstrunk} and ``topoelectrical'' circuits \cite{Segev}. Each of these physical implementations has its own advantages and disadvantages. Indisputably, condensed matter systems are the go-to platform for studying many-fermion phenomena, while ultracold atoms take the lead in the many-boson counterpart. For obvious reasons, classical (mechanical, electrical and optical) systems are particularly suitable for the preparation of analog one-body and, in the case of photonic lattices, even two-body physics in topologically non-trivial media.

Much of the interesting phenomenology of topologically non-trivial systems may be attributed to the bulk-edge correspondence \cite{ZhangReview}, which relates the existence of a concrete number of in-gap bound states attached to the edges or surfaces of the bulk samples to topological invariants of the bulk single-particle states. Differentiating these topological edge states from trivial Shockley waves \cite{Shockley} in the spectrum is easily done by making the edges infinitely sharp: the energy of a Shockley wave diverges as the potential at the edge is made stronger, while the energy of topological edge states remains finite and converges to a well-defined value for a sharp edge \cite{Hatsugai,Slager,Hugel,DuncanOhbergValiente}. What is not clear, however, is the fate of topological edge states when the edge is a quantum, dynamical entity. This is particularly simple to imagine in one spatial dimension, where a sharp edge can be modeled by a single immobile impurity. Upon allowing the heavy impurity to move quantum-mechanically, the edge modes lose their meaning as single-particle effects and, instead, a two-body problem of distinguishable particles emerges. The question of stability of the topological edge modes is equivalent to asking whether finite-energy two-body bound states exist in the limit of hard core, on-site particle-impurity interactions. Here, I consider a particle in the one-dimensional Su-Schrieffer-Heeger (SSH) model interacting with a mobile quantum impurity via a hard core on-site potential. The kinematics of the impurity is described by a standard Hubbard-type hopping Hamiltonian, which is topologically trivial. This may be regarded as the simplest non-trivial few-body problem in a topological setting that requires a fully non-perturbative treatment. By solving the two-body problem analytically, I find that not only are the edge modes robust to the mobility of the impurity, but also that the system forms a flat band of two-body bound states with energies identical to the energy of the original edge mode. The flat band allows for the construction of fully localised eigenstates, that is, a bound state of a particle with a heavy impurity localises the impurity. This flat band of topological bound states is robust as long as the spectrum of the two-body system remains gapped. This phenomenology is to be contrasted with studies of the two-body problem with identical particles \cite{ValienteKuesterSaenz,DiLiberto1,DiLiberto2,Qin1,Qin2}, for which two-body bound states of finite energy for hard core interactions are forbidden by the Bose-Fermi mapping theorem \cite{ValienteKuesterSaenz,DiLiberto1,Qin1,Qin2} and/or by the gapless nature of the two-body spectrum at the energies of the topological edge modes \cite{DiLiberto2}. Instead, interesting edge modes composed of doublons may exist in these systems \cite{DiLiberto1,DiLiberto2,Qin1,Qin2}. In a Hofstadter two-rung ladder, it has been proven, experimentally \cite{Greiner}, that two interacting atoms form chiral bound states due to super-exchange and therefore have finite energy. These, however, disappear into the scattering continuum for strong interactions. It is also worth noting that polarons in non-trivial topological systems have recently attracted some attention \cite{Grusdt1,Grusdt2,Bruun,Cui} and host a variety of interesting phenomena. 

The Hamiltonian of the system is given by
\begin{equation}
  H=H_{0}+H_{\mathrm{I}}+U\mathcal{V},\label{FullH}
\end{equation}
where $H_0$, $H_{\mathrm{I}}$ and $U\mathcal{V}$ are, respectively, the SSH Hamiltonian, the trivial impurity Hamiltonian and the particle-impurity interaction, given by
\begin{align}
  H_0 &= \sum_{j}\left[-t+(-1)^j\delta\right]c_{j+1}^{\dagger}c_j+\mathrm{H.c.}, \label{HSSH}\\
  H_{\mathrm{I}} &= -J\sum_{j}a_{j+1}^{\dagger}a_j+\mathrm{H.c.},\label{HI}\\
  \mathcal{V} &= \sum_j c_j^{\dagger}c_ja_{j}^{\dagger}a_{j}.
\end{align}
Above, $c_j$ and $a_j$ ($c_j^{\dagger}$ and $a_{j}^{\dagger}$) are annihilation (creation) operators at site $j$ for either fermions or bosons, as we shall work in the two-body sector (one particle in the SSH chain and one impurity). All pairs of different creation and annihilation operators commute with each other, i.e. $[a_j,c_{j'}]=[a_j,c_{j'}^{\dagger}]=[a_j^{\dagger},c_{j'}]=[a_j^{\dagger},c_{j'}^{\dagger}]=0$ $\forall j,j'$. For concreteness and without loss of generality I shall consider $J,t,\delta>0$ unless otherwise stated, with $\delta<t$.

The single-particle eigenstates of $H_0$ and $H_{\mathrm{I}}$ are well known, and I review them here briefly. I write them as $\ket{\psi^{(0)}}=\sum_{j}\psi^{(0)}(j)c_j^{\dagger}\ket{0}$ and $\ket{\psi^{(\mathrm{I})}}=\sum_{j}\psi^{(\mathrm{I})}(j)a_j^{\dagger}\ket{0}$, with $\ket{0}$ the particle vaccuum. The states of the SSH model take the form dictated by Bloch's theorem, i.e. $\psi^{(0)}(j)= \phi^{(0)}(j)\exp{(ikj)}$, with $\phi^{(0)}(j+2)=\phi^{(0)}(j)$. The Bloch function therefore takes on only two values, related by $\phi^{(0)}(1)=\exp{(i\theta_k)}\phi^{(0)}(0)$, with
\begin{align}
  e^{i\theta_k}&=\pm \frac{\left|\xi_{k}\right|}{\xi_k},\\
  \xi_k &\equiv -2t\cos k + 2i\delta \sin k,\label{xi}
\end{align}
where the $-$ and $+$ signs refer, respectively, to the first and second Bloch bands. Their associated eigenenergies are given by
\begin{equation}
  E_{\pm}^{(0)}(k) =\pm |\xi_k|, \hspace{0.1cm} k\in (-\pi/2,\pi/2].\label{energySSH}
\end{equation}
The eigenstates of the impurity Hamiltonian $H_{\mathrm{I}}$ are simply plane waves $\psi^{(\mathrm{I})}(j) = \exp(iqj)$, with energy dispersion
\begin{equation}
  E^{(\mathrm{I})}(q)=-2J\cos(q), \hspace{0.1cm} q\in (-\pi,\pi].\label{energyImpurity}
\end{equation}
With boundaries in the system, the SSH model admits topological edge modes, which correspond to setting exponentially decaying (growing) Bloch waves if the particle is attached to the right (left) of an infinite wall. The ansatz takes the form $\psi^{(0)}(j)=\phi_{\alpha}(j)\alpha^j$, with $\phi_{\alpha}(j+2)=\phi_{\alpha}(j)$. Assume now that there is an infinite wall at $j=x_0$, with no other boundaries. Introducing $\psi^{(0)}(j)$ into the stationary Schr{\"o}dinger equation $H_{0}\ket{\psi^{(0)}}=E\ket{\psi^{(0)}}$, it is simple to show that $E=0$, $\phi_{\alpha}(x_0)=0$ and
\begin{equation}
\alpha = \sqrt{\frac{\delta+t}{\delta-t}}.\label{alphaSSH}
\end{equation}   
From Eq.~(\ref{alphaSSH}) a topological phase transition is inferred at $\delta/t=0$, where the bulk spectrum, Eq.~(\ref{energySSH}), becomes gapless. The two phases can be characterised by their change in Zak's phase \cite{Cooper}, or by means of the nature of their edge states. For $x_0$ even (odd), if $\delta/t<0$, then $|\alpha|<1$ ($|\alpha|>1$) and therefore the edge state is bound to the right (left) of $x_0$. For $\delta/t>0$, the exact opposite is true while, for $\delta=0$, the system is topologically trivial (no edge state) and behaves identically to the mobile impurity with Hamiltonian (\ref{HI}). 

After the single-particle prelude, it is instructive to begin with an immobile impurity, by setting $J=0$ in Eq.~(\ref{HI}). Without interactions, and forgetting about the SSH model for now, the impurity can occupy $N$ different sites in an $N$-site lattice, and the localised states $a_j^{\dagger}\ket{0}$ are all eigenstates with zero energy. I set now the interactions to hard core ($U\to \infty$), while retaining $J=0$. A particle in the SSH model, Eq.~(\ref{HSSH}), coupled to the static impurity can form exactly $m=N$ bound states assuming $N$ is even and periodic boundary conditions (with other conditions on the boundary and commensurability, $m\ne N$ but $m/N\to 1$ as $N\to \infty$), all with zero energy. This is of course not surprising. By continuity arguments, at least some of these bound states -- which are a generalisation of topological edge states --  should remain stable for a mobile impurity with $J\ne 0$, even in the hard core limit. The important remaining questions are how these topological bound states disperse as the impurity is allowed to travel along the lattice, and what the criteria for their stability consist of. 

\begin{figure}[t]
\includegraphics[width=0.45\textwidth]{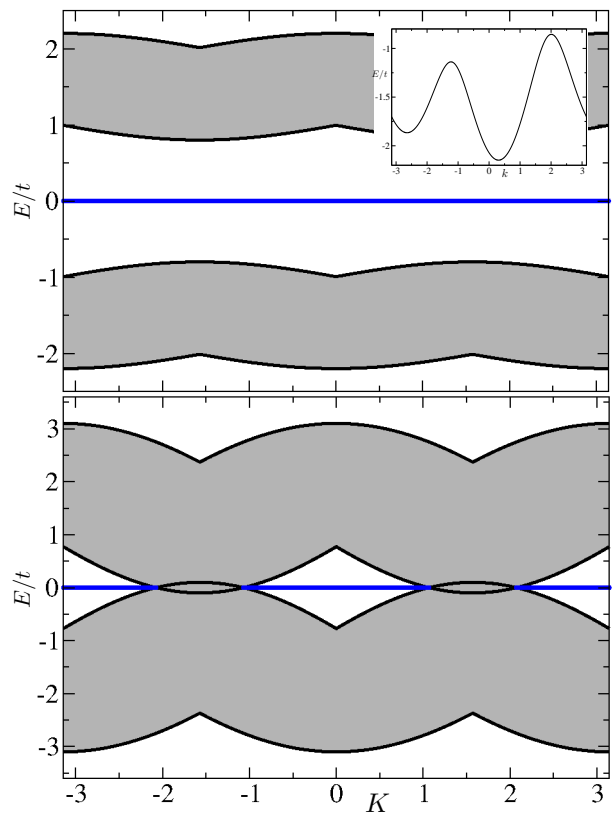}
\caption{Upper panel: spectrum of Hamiltonian (\ref{FullH}) with $J/t=1/10$, $\delta/t=1/2$ and $t/U=0$. Shaded areas correspond to the scattering continuum, blue dots to the bound states and black solid lines to the lowest and highest continuum energies for each of the two bands. Inset: lowest two-body energy band $E(K,k)$ for fixed $K=\pi/4$ as a function of the relative quasi-momentum $k$. Lower panel: same as upper panel but with $J/t=0.55$, $\delta/t=1/2$ and $t/U=0$.}
\label{fig:spectrum}
\end{figure}

To solve the interacting particle-impurity problem, I introduce two-body wave functions $\ket{\Psi}$, as
\begin{equation}
  \ket{\Psi}=\sum_{x,y}\Psi(x,y)c_x^{\dagger}a_y^{\dagger}\ket{0},
\end{equation}
where I shall consider an infinitely long lattice. Finite lattices can be investigated by properly choosing combinations of infinite-size eigenfunctions, including exponentially growing, non-normalisable solutions \footnote{Note that this is in general the case for standard bound states. The bound state wave functions $\psi(x)$ of a single particle in a finite box with well-defined but arbitrary boundary conditions behave, away from the range of the binding potential, as $\psi(x)\propto e^{-\lambda x}+Be^{\lambda x}$, with real $\lambda$.}, with identical eigenenergies. The stationary Schr{\"o}dinger equation $H\ket{\Psi}=E\ket{\Psi}$, with $H$ given in Eq.~(\ref{FullH}), is written as
\begin{align}
  &\left[-t+(-1)^{x+1}\delta\right]\Psi(x-1,y)+\left[-t+(-1)^x\delta\right]\Psi(x+1,y)\nonumber\\
  &-J\left[\Psi(x,y+1)+\Psi(x,y-1)\right]+U\delta_{x,y}\Psi(x,y)\nonumber \\
  &=E\Psi(x,y),\label{Recurrence2Body}
\end{align}  
where $\delta_{x,y}$ is a Kronecker delta. For $\delta=0$ (topologically trivial), Eq.~(\ref{Recurrence2Body}) reduces to the two-body problem with unequal masses on a homogeneous lattice with nearest-neighbour hopping, which admits a simple exact solution \cite{ValientePRA,PiilMolmer}. As I shall show now, Eq.~(\ref{Recurrence2Body}) can be solved analytically for arbitrary values of $\delta$, albeit with much more complicated solution. To see this, note that $H$, Eq.~(\ref{FullH}), conserves total quasi-momentum $K=k_1+k_2$ ($\mathrm {mod}$ $2\pi$). The usual separation of the wave function into coordinates $R=(x+y)/2$ and $z=x-y$ as $\Psi(x,y)\sim \exp(iKR)g(z)$, however, is {\it not} correct. This is due to the non-trivial periodicity of the SSH Hamiltonian. Instead, the following separation is necessary
\begin{align}
  \Psi(x,y)&=e^{iKR}\psi_K(x,z),\label{Separation1}\\
  \psi_K(x+2,z)&=\psi_K(x,z).\label{Separation11}
\end{align}
Notice that Eq.~(\ref{Separation1}), together with the periodicity condition (\ref{Separation11}), holds, too, for other periodic potentials with period $\tau>2$ by setting $\psi_K(x+\tau,z)=\psi_K(x,z)$. Because of periodicity, Eq.~(\ref{Separation11}), there are only two functions of the relative coordinate that need to be determined, namely $\psi_K(0,z)\equiv \Phi_K^{0}(z)$ and $\psi_K(1,z)\equiv \Phi_K^{1}(z)$. Therefore, upon substitution of Eq.~(\ref{Separation1}) into Eq.~(\ref{Recurrence2Body}), this reduces to a set of two coupled difference equations
\begin{align}
  &-(t+\delta)e^{-iK/2}\Phi_K^{1}(z+1)-(t-\delta)e^{iK/2}\Phi_K^{1}(z-1)\nonumber\\
  &-J\left[e^{iK/2}\Phi_K^{0}(z+1)+e^{-iK/2}\Phi_K^{0}(z-1)\right]+U\delta_{z,0}\Phi_K^{0}(z)\nonumber\\
  &=E\Phi_K^0(z),\label{Recurrence3}\\
  &-(t-\delta)e^{-iK/2}\Phi_K^{0}(z+1)-(t+\delta)e^{iK/2}\Phi_K^{0}(z-1)\nonumber\\
  &-J\left[e^{iK/2}\Phi_K^{1}(z+1)+e^{-iK/2}\Phi_K^{1}(z-1)\right]+U\delta_{z,0}\Phi_K^{1}(z)\nonumber\\
  &=E\Phi_K^{1}(z).\label{Recurrence4}
\end{align}
For $U=0$, the functions $\Phi_K^{j}(z)$ can be written as Bloch states associated with period $\tau=2$ (in $z$) as well. Upon turning on interactions, the following asymptotic condition holds for scattering states where the impurity is incident from the right
\begin{align}
  \Phi_K^{j}(z)&\propto \left[u_{K,k}^j(z)e^{ikz}+r_{K,k}u_{K,k'}^{j}(z)e^{ik'z}\right]\bar{\theta}(-z)\nonumber\\
  &+t_{K,k}u_{K,k}^j(z)e^{ikz}\theta(z),\hspace{0.1cm} |z|\to \infty,\label{ScatteringAsymptotics1}
\end{align}
Above, $u_{K,k}^j(z+2)=u_{K,k}^j(z)$ is the periodic Bloch function associated with relative quasi-momentum $k$, such that the group velocity $\mathrm{d}E(K,k)/\mathrm{d}k>0$, with $E(K,k)$ the non-interacting two-body energy dispersion (I have obviated the band indices for simplicity of notation), and $k'$ is such that $E(K,k')=E(K,k)$ \footnote{For indistinguishable particles, $k' = -k$}. In Eq.~(\ref{ScatteringAsymptotics1}), $r_{K,k}$ and $t_{K,k}$ constitute two-body generalisations of transmission and reflection coefficients, and $\theta(z)$ ($\bar{\theta}(z)$) is the Heaviside step function being zero (unity) at $z=0$. The asymptotic form of the scattering states, Eq.~(\ref{ScatteringAsymptotics1}), is valid only for total momenta $K$ and energies $E(K,k)$ for which no four-wave mixing exits, since for non-trivial dispersions such as the one of the SSH model, it is possible to find $K$, $k_1$, $k_2$, $k_3$ and $k_4$ such that $E(K,k_1)=E(K,k_2)=E(K,k_3)=E(K,k_4)$, satisfying $k_i\ne k_j$ for $i\ne j$, as seen in the inset of Fig.~\ref{fig:spectrum}. In such cases, Eq.~(\ref{ScatteringAsymptotics1}) needs to be modified by including and extra term \cite{ValienteKuesterSaenz}. For the impurity incident from the left, the asymptotic expression of the scattering states is completely analogous to Eq.~(\ref{ScatteringAsymptotics1}), with $k$ corresponding to negative group velocity, and besides the possible modification due to four wave mixing. The scattering states can be obtained exactly for all values of the interaction $U$ by either including the degenerate state when four-wave mixing occurs, or a virtual bound state otherwise, in an exactly analogous fashion to the case of identical particles in the ionic Hubbard model \cite{ValienteKuesterSaenz}. Since interactions do not modify the continuous spectrum and the goal here is to study topological bound states, I shall not discuss scattering states any further.

I now turn my attention to the particle-impurity bound states. If finite-energy bound states survive the hard core ($U\to \infty$) limit, then these are necessarily associated with the non-trivial topology of the SSH model, and represent the generalisation of edge states to the case of a mobile impurity. The hard core limit is here not simpler than the general case, due to the distinguishability of the particle and impurity, in clear contrast with two identical particles in the SSH model \cite{ValienteKuesterSaenz}. This is to say that, even for $U\to \infty$, the many-body problem is not integrable. I shall prove that, for $U\to \infty$, there are infinitely many zero-energy states regardless of whether zero energy is in the scattering continuum (i.e. the non-interacting spectrum) or not. This fact implies that if the zero of energy is not in the scattering continuum, every state with vanishing energy must correspond to a bound state. The non-interacting spectrum contains the zero of energy only if $J\ge \delta$. Exactly at $J=\delta$, this happens at the points where the two two-body bands touch, namely for $K=\pm \pi/2$ ($k=\mp \pi/4$). As we increase $J$, the bands start to overlap in larger intervals of total momentum $K$ until the gap is closed at all momenta when $J\ge t$. In order for zero-energy states to exist, a necessary but not sufficient condition is that the Schr{\"o}dinger equation, Eqs.~(\ref{Recurrence3},\ref{Recurrence4}), admits four asymptotic ($z\to\pm\infty$) solutions of the form $\Phi_K^j(z) \propto \alpha_K^{z}$ with $|\alpha_K|\ne 1$. Two of these solutions need to fulfill $|\alpha_K|<1$, while the other two must satisfy $|\alpha_K|>1$. Clearly, the first condition implies the latter, since if a wave function corresponding to $\alpha_K$ is a solution to Eqs.~(\ref{Recurrence3},\ref{Recurrence4}) for $z>1$, so is the one corresponding to $\alpha_K^{-1}$. That this needs to be the case in order for zero energy states to exist is easily shown in two steps: (i) There is no solution of the form $\Phi_0(z)=0$ ($\Phi_1(z)=0$) for all $z$ while leaving $\Phi_1(z)\ne 0$ ($\Phi_0(z)\ne 0$) \footnote{This is the form of the edge state, i.e. for $J=0$, see Ref.~\cite{DuncanOhbergValiente}}, which is immediately checked by inserting that type of solution into Eqs.~(\ref{Recurrence3},\ref{Recurrence4}) at $z=0$ and $z=1$; (ii) given (i), in order for the two-body wave function to vanish at $z=0$ it is necessary to superimpose two asymptotic bound state wave functions corresponding to $\alpha_K$ and $\alpha_K'$, with $\alpha_K\ne \alpha_K',\alpha_K'^{-1}$, respectively. Since the interaction in Eq.~(\ref{FullH}) has zero range, the coupled set of recurrence relations (\ref{Recurrence3},\ref{Recurrence4}) only depends on the asymptotic solutions for $z\ge 2$. Defining an asymptotic wave function as $\tilde{\Phi}_K^j(z)=A_j\alpha_K^z$, and inserting this into the recurrence relations (\ref{Recurrence3},\ref{Recurrence4}) for $z>1$ and $E=0$, I obtain
\begin{align}
  \frac{A_0}{A_1}&=\frac{J\left[\alpha_K^2e^{iK/2}+e^{-iK/2}\right]}{(-t+\delta)\alpha_K^2e^{-iK/2}-(t+\delta)e^{iK/2}}.\label{A0A1}\\
  \alpha_{K}&=(-1)^{s_1}\sqrt{\frac{1}{2a_K}\left[-b_K+(-1)^{s_2}\sqrt{b_K^2-4|a_K|^2}\right]},\label{alpha}
\end{align}
where $s_i=0$, $1$ ($i=1$, $2$) give the four roots for $\alpha_K$, and where
\begin{align}
  a_K&=J^2e^{iK}-(t^2-\delta^2)e^{-iK}\\
  b_K&=2\left[J^2-(t^2+\delta^2)\right].
\end{align}
The values of $\alpha_K$, Eq.~(\ref{alpha}), all satisfy $|\alpha_K|\ne 1$ if $J<\delta$. For $\delta\le J< t$, they satisfy this condition for values of the total momentum $K$ for which there is a gap, while for $J\ge t$ it is not satisfied for any $K$, as expected. Importantly, if $\alpha_K$ is a solution, so is $-\alpha_K$, see Eq.~(\ref{alpha}). This implies that $A_0/A_1$, Eq.~(\ref{A0A1}), which depends on $\alpha_K$ only as $\alpha_K^2$, is independent of the value of $s_1$ in Eq.~(\ref{alpha}). Therefore, zero-energy bound states with hard core interactions can be constructed and take the form
\begin{equation}
  \Phi_K^j(z) = A_j\alpha_K^{|z|}\left[1+(-1)^{z}B_K\right]\left[\bar{\theta}(z)+\gamma_K\theta(-z)\right],\label{BoundStateWF}
\end{equation}
where $\gamma_K$ is a constant yet to be determined, while $B_K=-1$ in virtue of the hard core boundary conditions $\Phi_K^j(0)=0$, $j=0,1$.

I proceed now to the calculation of $\gamma_K$ in Eq.~(\ref{BoundStateWF}), which is slightly subtle, since $\lim_{U\to \infty}U\Phi_K^j(0)$, Eq.~(\ref{Recurrence3},\ref{Recurrence4}), is non-zero and needs to be calculated. I now allow for $U$ to be large but finite. Therefore, $B_K=B_K(U)\ne -1$ in Eq.~(\ref{BoundStateWF}), and is unknown. Since $\Phi_K^j(0)$ is only required to first order in $1/U$ in perturbation theory, I write $B_K=-1+\beta_K/U$, while the perturbative expansion of $\gamma_K$ in Eq.~(\ref{BoundStateWF}) is irrelevant. It is immediate to find that
\begin{equation}
\lim_{U\to \infty}U\Phi_K^j(0)= \beta_KA_j.\label{lambdaj}
\end{equation}
Introducing Eq.~(\ref{lambdaj}) into the recurrence relations (\ref{Recurrence3},\ref{Recurrence4}), simple algebraic manipulations yield
\begin{align}
  \gamma_K&=\frac{-(t-\delta)A_0/A_1+(t+\delta)A_1/A_0}{(t-\delta)A_1/A_0+(t+\delta)A_0/A_1}e^{-iK},\label{gamma}\\
  \beta_K&=\left[-(t-\delta)\frac{A_1}{A_0}\gamma_Ke^{iK/2}+(t+\delta)\frac{A_1}{A_0}e^{-iK/2}\right.\nonumber\\
    &\left.\hspace{0.5cm}+Je^{iK/2}+J\gamma_Ke^{-iK/2}\right]\alpha_K.\label{beta}
\end{align}
Above, $A_0/A_1$ is given by Eq.~(\ref{A0A1}), with $\alpha_K$ given by either of its values in Eq.~(\ref{alpha}) with $|\alpha_K|<1$. This concludes the derivation of the topological bound states in the hard core limit. For the sake of clarity, I summarize their properties: (i) For all $K$ such that there is a gap in the middle of the spectrum, there is one bound state with energy $E=0$; (ii) the bound state eigenfunctions are given by $\Psi(x,y)=\exp(iKR)\Phi_K^x(z)$, with $K$ the total quasi-momentum, $R=(x+y)/2$ and $z=x-y$, and $\Phi_K^x(z)$ given by Eq.~(\ref{BoundStateWF}), with (iii) $\alpha_K$ in Eq.~(\ref{alpha}), $A_j$ defined by Eq.~(\ref{A0A1}) (up to an arbitrary normalisation constant), $B_K=-1$ and $\gamma_K$ defined in Eq.~(\ref{gamma}). The spectrum of the two-body system is plotted in Fig.~\ref{fig:spectrum}, where it is clear that for a fully gapped continuum (upper panel, $J<\delta$) there is one bound state per total momentum $K$, while for a partial gap bound states exist only in-gap (lower panel, $\delta\le J<t$) and, obviously, none exist without a gap (not shown, $J\ge t$).

\begin{figure}[t]
\includegraphics[width=0.45\textwidth]{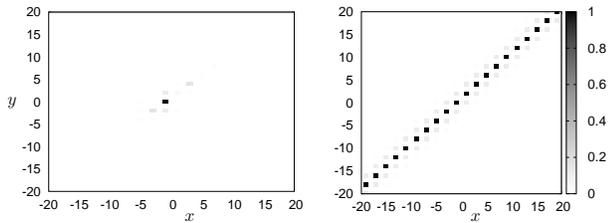}
\caption{Probability distribution $|\Psi(x,y)|^2$ of zero-energy bound states with $J/t=1/10$ and $\delta/t=1/2$. Left panel: bound state obtained using Eq.~(\ref{superposition}) with $f(K)=\cos(K)$. Right panel: bound state with well-defined total momentum $K=0$. Values of $|\Psi(x,y)|^2$ are normalised to the peak.}
\label{fig:WF}
\end{figure}

The most interesting consequence of the persistence of a flat band of topological bound states is the fact that fully localised eigenstates can be constructed using linear combinations of bound states for all $K$ \cite{Huber,Pal,ValienteZinnerFlatBand}. It is possible to construct localised eigenstates as
\begin{equation}
  \Psi(x,y)=\int_{-\pi}^{\pi}\mathrm{d}Kf(K)e^{iKR}\Phi_K^x(z),\label{superposition}
\end{equation}
where $f(K)$ is an arbitrary function of the total quasi-momentum. In Fig.~\ref{fig:WF}, I show the probability distributions of a bound state obtained using Eq.~(\ref{superposition}) with $f(K)=\cos(K)$ (left panel), and of a translationally invariant bound state with $K=0$ (right panel). Clearly, the eigenstate in the left panel of Fig.~\ref{fig:WF} is localised and normalisable.   

The results here presented can be observed with current state-of-the-art experiments, whether in ultracold atomic settings \cite{Greiner,Bloch}, or classical-optical systems, where a one-dimensional two-body problem for distinguishable particles can be simulated by means of a two-dimensional photonic lattice \cite{b1,b2,b3,b4,b5,b6,b7,b8,Mukherjee}. In particular, the SSH model may be simulated along the horizontal axis \cite{BlancoRedondo}, the standard, trivial impurity Hamiltonian in the vertical axis \cite{b5,Mukherjee}, while the hard-core two-body interaction can be engineered by a large shift in the propagation constant along the diagonal \cite{b5,Mukherjee}. The presence of a flat band of topological bound states in a photonic lattice is easily inferred by exciting a wave guide at position $(x,y,z)=(x,x\pm 1,0)$ \footnote{The direction $z$ here is analogous to time in quantum mechanics.} and measuring the output at non-zero $z$ (the end of the sample), which will remain localised around its initial position.

In summary, I have considered a particle in the one-dimensional SSH model interacting strongly with a heavy mobile impurity and shown that the system exhibits a flat band of zero energy topological bound states as long as the two-body continuum is gapped. This implies that an impurity can be put to rest by the lighter, topologically non-trivial particle. This phenomenon is possible due to the fact that, in one dimension, impurities have the same (zero) spatial dimension as the sample's edge. It would be interesting to extend these results to higher dimensions where, instead of a mobile impurity, one may consider a cold, continuous edge or surface with quantised rotational and vibrational modes, and ask whether these modes can be suppressed by means of topological bound states. This may be possible using nanoparticle topological insulators \cite{d1,d2,d3}.

\acknowledgements I am grateful to Hong Yao and Tian-Shu Deng for useful discussions.

\bibliographystyle{unsrt}

\begin{thebibliography}{99}
\bibitem{Bernevig} B.~A. Bernevig, T.~L. Hughes and S.~C. Zhang, Science {\bf 314}, 1757 (2006).

\bibitem{Kane} C.~L. Kane and E.~J. Mele, Phys. Rev. Lett. {\bf 95}, 226801 (2005).

\bibitem{Koenig} M. K{\"o}nig, S. Wiedmann, C. Br{\"u}ne, A. Roth, H. Buhmann, L.~W. Molenkamp, X.-~L. Qi and S.-~C. Zhang, Science {\bf 318}, 776 (2007).  

\bibitem{Hall1} K. von Klitzing, G. Dorda and M. Pepper, Phys. Rev. Lett. {\bf 45}, 494 (1980).

\bibitem{Hall2} R.~B. Laughling, Phys. Rev. B {\bf 23}, 5632 (1981).

\bibitem{Hall3} D.~J. Thouless, M. Kohmoto, M.~P. Nightingale and M. den Nijs, Phys. Rev. Lett. {\bf 49}, 405 (1982).
  
\bibitem{Fractional1} D.~C. Tsui, H.~L. Stormer and A.~C. Gossard, Phys. Rev. Lett. {\bf 48}, 1559 (1982).

\bibitem{Fractional2} R.~B. Laughling, Phys. Rev. Lett. {\bf 50}, 1395 (1983).

\bibitem{Cooper} N.~R. Cooper, J. Dalibard and I.~B. Spielman, Rev. Mod. Phys. {\bf 91}, 015005 (2019).

\bibitem{reviewphotonics} T. Ozawa, H.~M. Price, A. Amo, N. Goldman, M. Hafezi, L. Lu, M. Rechtsman, D. Schuster, J. Simon, O. Zilberberg and I. Carusotto, Rev. Mod. Phys. {\bf 91}, 015006 (2019).

\bibitem{HuberSusstrunk} R. S{\"u}sstrunk and S.~D. Huber, Science {\bf 349}, 47 (2015).
  
\bibitem{Segev} N.~A. Olekhno, E.~I. Kretov, A.~A. Stepanenko, D.~S. Filonov, V.~V. Yaroshenko, B. Cappello, L. Matekovits and M.~A. Gorlach, e-print arXiv:1907.01016v1 .  
  
%\bibitem{a1} A. Roth, C. Br{\"u}ne, H. Buhmann, L.~W. Molenkamp, J. Maciejko, X.~-L. Qi and S.~-C. Zhang, Science {\bf 325}, 294 (2009).

%\bibitem{a2} K. Suzuki, Y. Harada, K. Onomitsu and K. Muraki, Phys. Rev. B {\bf 87}, 235311 (2013).

%\bibitem{a3} G. Grabecki, J. Wr{\'o}bel, M. Czapkiewicz, L. Cywi{\'n}ski, S. Gieraltowska, E. Guziewicz, M. Zholudev, V. Gavrilenko, N.~N. Mikhailov, S.~A. Dvoretski, F. Teppe, W. Knap and T. Dietl, Phys. Rev. B {\bf 88}, 165309 (2013).

%\bibitem{a4} L. Du, I. Knez, G. Sullivan and R.~R. Du, Phys. Rev. Lett. {\bf 114}, 096802 (2015).

%\bibitem{a5} M. K{\"o}nig, M. Baenninger, A.~G.~F. Garcia, N. Harjee, B.~L. Pruitt, C. Ames, P. Leubner, C. Br{\"u}ne, H. Buhmann, L.~W. Molenkamp and D. Goldhbaer-Gordon, Phys. Rev. X {\bf 3}, 021003 (2013).
  
%\bibitem{Moore} C. Xu and J.~E. Moore, Phys. Rev. B {\bf 73}, 045322 (2006).

%\bibitem{c} S. Wu, V. Fatemi, Q.~D. Gibson, K. Watanabe, T. Taniguchi, R.~J. Cava and P. Jarillo-Herrero, Science {\bf 359}, 76 (2018).

%\bibitem{Novelli} P. Novelli, F. Taddei, A.~K. Geim and M. Polini, Phys. Rev. Lett. {\bf 122}, 016601 (2019).

\bibitem{ZhangReview} X.~-L. Qi and S.~-C. Zhang, Rev. Mod. Phys. {\bf 83}, 1057 (2011).

\bibitem{Shockley} W. Shockley, Phys. Rev. {\bf 56}, 317 (1939).

\bibitem{Hatsugai} Y. Hatsugai, Phys. Rev. Lett. {\bf 71}, 3697 (1993).

\bibitem{Slager} R.~-J. Slager, L. Rademaker, J. Zaanen and L. Balents, Phys. Rev. B {\bf 92}, 085126 (2015).
  
\bibitem{Hugel} D. H{\"u}gel and B. Paredes, Phys. Rev. A {\bf 89}, 023619 (2014).
  
\bibitem{DuncanOhbergValiente} C.~W. Duncan, P. {\"O}hberg and M. Valiente, Phys. Rev. B {\bf 97}, 195439 (2018).  

\bibitem{ValienteKuesterSaenz} M. Valiente, M. K{\"u}ster and A. Saenz, EPL {\bf 92}, 10001 (2010).

\bibitem{DiLiberto1} M. Di Liberto, A. Recati, I. Carusotto and C. Menotti, Phys. Rev. A {\bf 94}, 062704 (2016).

\bibitem{DiLiberto2} G. Salerno, M. Di Liberto, C. Menotti and I. Carusotto, Phys. Rev. A {\bf 97}, 013637 (2018).

\bibitem{Qin1} X. Qin, F. Mei, Y. Ke, L. Zhang and C. Lee, Phys. Rev. B {\bf 96} 195134 (2017). 

\bibitem{Qin2} X. Qin, F. Mei, Y. Ke, L. Zhang and C. Lee, New J. Phys. {\bf 20}, 013003 (2018).

\bibitem{Greiner} M.~E. Tai, A. Lukin, M. Rispoli, R. Schittko, T. Menke, D. Borgnia, P.~M. Preiss, F. Grusdt, A.~M. Kaufman and M. Greiner, Nature {\bf 546}, 519 (2017).
  
\bibitem{Grusdt1} F. Grusdt, N.~Y. Yao, D. Abanin, M. Fleischhauer and E. Demler, Nature Comms. {\bf 7}, 11994 (2016).

\bibitem{Grusdt2} F. Grusdt, N.~Y. Yao and E. Demler, e-print arXiv:1904.00220 .
  
\bibitem{Bruun} A. Camacho-Guardian, N. Goldman, P. Massignan and G.~M. Bruun, Phys. Rev. B {\bf 99}, 081105 (2019).

\bibitem{Cui} F. Qin, X. Cui and W. Yi, Phys. Rev. A {\bf 99}, 033613 (2019).
  
\bibitem{ValientePRA} M. Valiente, Phys. Rev. A {\bf 81}, 042102 (2010).

\bibitem{PiilMolmer} R.~T. Piil, N. Nygaard and K. M{\o}lmer, Phys. Rev. A {\bf 78}, 033611 (2008).

\bibitem{Huber} S.~D. Huber and E. Altman, Phys. Rev. B {\bf 82}, 184502 (2010).

\bibitem{Pal} B. Pal, Phys. Rev. B {\bf 98}, 245116 (2018).
  
\bibitem{ValienteZinnerFlatBand} M. Valiente and N.~T. Zinner, J. Phys. B: At. Mol. Opt. Phys. {\bf 50}, 064004 (2017).

\bibitem{Bloch} M. Atala, M. Aidelsburger, J.~T. Barreiro, D. Abanin, T. Kitagawa, E. Demler and I. Bloch, Nature Phys. {\bf 9}, 795 (2013).

\bibitem{b1} S. Longhi, J. Phys. B: At. Mol. Opt. Phys. {\bf 44}, 051001 (2011).

\bibitem{b2} D.~O. Krimer and R. Khomeriki, Phys. Rev. A {\bf 84}, 041807 (2011).

\bibitem{b3} S. Longhi, Opt. Lett. {\bf 36}, 3248 (2011).

\bibitem{b4} S. Longhi and G. Della Valle, Phys. Rev. A {\bf 86}, 042104 (2012).

\bibitem{b5} G. Corrielli, A. Crespi, G. Della Valle, S. Longhi and R. Osellame, Nature Comms. {\bf 4}, 1555 (2013).

\bibitem{b6} A. Rai, C. Lee, C. Noh and D.~G. Angelakis, Sci. Rep. {\bf 5}, 8438 (2015).

\bibitem{b7} K. Noba, Phys. Rev. B {\bf 67}, 153102 (2003).

\bibitem{b8} S. Longhi and G. Della Valle, Opt. Lett. {\bf 36}, 4743 (2011).
  
\bibitem{Mukherjee} S. Mukherjee, M. Valiente, N. Goldman, A. Spracklen, E. Andersson, P. {\"O}hberg and R.~R. Thomson, Phys. Rev. A {\bf 94}, 053853 (2016).

\bibitem{BlancoRedondo} A. Blanco-Redondo, B. Bell, D. Oren, B.~J. Eggleton and M. Segev, Science {\bf 362}, 568 (2018).

\bibitem{d1} G. Siroki, D.~K.~K. Lee, P.~D. Haynes and V. Giannini, Nature Comms. {\bf 7}, 12375 (2016).

\bibitem{d2} K.~-I. Imura, Y. Yoshimura, Y. Takane and T. Fukui, Phys. Rev. B {\bf 86}, 235119 (2012).

\bibitem{d3} D.~-H. Lee, Phys. Rev. Lett. {\bf 103}, 196804 (2009).
  
\end{thebibliography}

\end{document}